\documentclass[useAMS,usenatbib]{mn2e}
\usepackage{natbib}
\usepackage{epsfig}
\usepackage{rotating}

\def\lesssim{\,\lower2truept\hbox{${<\atop\hbox{\raise4truept\hbox{$\sim$}}}$}\,}
\def\gtrsim{\,\lower2truept\hbox{${>\atop\hbox{\raise4truept\hbox{$\sim$}}}$}\,}

\title[The PACO multifrequency analysis]{The {\it Planck}-ATCA Co-eval Observations (PACO) project: analysis of radio source properties between 5 and 217\,GHz.}
\author[M. Massardi et al.]{
\parbox[t]{\textwidth}
{Marcella Massardi$^{1}$\thanks{E-mail: massardi@ira.inaf.it}, Anna Bonaldi$^{2}$, Laura Bonavera$^{3}$, Gianfranco De Zotti$^{4,5}$, Marcos Lopez-Caniego$^{3}$, Vincenzo Galluzzi$^{1,6}$}
\vspace*{8pt} \\
$^{1}$INAF, Osservatorio di Radioastronomia, Via Gobetti 101, I-40129, Bologna\\
$^{2}$Jodrell Bank Centre for Astrophysics, School of Physics and Astronomy,\\ The University of Manchester, Oxford Road, Manchester M13 9PL, U.K.\\
$^{3}$European Space Agency, ESAC, Planck Science Office, Camino bajo del Castillo, s/n,\\ Urbanizaci\'{o}n Villafranca del Castillo, Villanueva de la Ca\~{n}ada, Madrid, Spain\\
$^{4}$INAF, Osservatorio Astronomico di Padova, Vicolo dell'Osservatorio 5, I-35122 Padova, Italy\\
$^{5}$SISSA, via Bonomea 265, I 34136 Trieste, Italy,\\
$^{5}$Dipartimento di Fisica e Astronomia, Universit{\`a} di Bologna, via Ranzani 1, I-40126 Bologna, Italy}
\begin{document}

\date{}

\pagerange{\pageref{firstpage}--\pageref{lastpage}} \pubyear{2010}

\maketitle

\label{firstpage}

\begin{abstract}
The {\it Planck}-ATCA Co-eval Observations (PACO) project has yielded observations of 464 sources with the Australia Telescope Compact Array (ATCA) between 4.5 and 40\,GHz. The main purpose of the project was to investigate the spectral properties of mm-selected radio sources at frequencies below and overlapping with the ESA's {\it Planck} satellite frequency bands, minimizing the variability effects by observing almost simultaneously with the first two {\it Planck} all-sky surveys. In this paper we present the whole catalogue of observations in total intensity. By comparing PACO with the various measures of {\it Planck} Catalog of Compact Sources (PCCS) flux densities we found the best consistency with the PCCS ``detection pipeline'' photometry (DETFLUX) that we used to investigate the spectral properties of sources from 5 to 217\,GHz. Of our sources, 91\,\% have remarkably smooth spectrum, well described by a double power law over the full range. This suggests a single emitting region, at variance with the notion that ``flat'' spectra result from the superposition of the emissions from different compact regions, self absorbed up to different frequencies. Most of the objects show a spectral steepening above $\simeq 30\,$GHz, consistent with synchrotron emission becoming optically thin. Thus, the classical dichotomy between flat-spectrum/compact and steep-spectrum/extended radio sources, well established at cm wavelengths, breaks down at mm wavelengths. The mm-wave spectra do not show indications of the spectral break expected as the effect of ``electron ageing'', suggesting young source ages.
\end{abstract}
\begin{keywords}
 galaxies: active -- radio continuum: galaxies -- catalogues.
\end{keywords}

\section{Introduction}


An important by-product of the {\it Planck} mission (Planck Collaboration 2011) are blind all-sky surveys of extragalactic sources over a broad frequency range, from 33 to 857\,GHz. These surveys have offered a unique opportunity to carry out an unbiased investigation of the spectral properties of radio sources in a poorly explored frequency range, partially unaccessible from the ground.  The {\it Planck} Catalog of Compact Sources (PCCS, Planck Collaboration XXVIII 2014) collects positions and flux densities of compact sources found in each of the nine {\it Planck} frequency maps for the nominal mission. It contains from several hundreds to thousands of extragalactic point sources (depending on frequency) and represents a major {\it Planck} contribution to multi-frequency studies of compact or point-like sources.
Ground-based observations can extend the spectral coverage at lower frequencies by orders of magnitude. However most extragalactic sources selected at high radio frequencies are strongly variable. This causes a major difficulty in interpreting the Spectral Energy Distribution (SEDs) obtained combining data from different experiments which are generally non-simultaneous.

To overcome this problem, the primary purpose of the {\it Planck}-ATCA Co-eval Observations (PACO) project was to carry out observations with the Australia Telescope Compact Array (ATCA) of sources potentially detectable by {\it Planck}, almost simultaneously with its observations and at frequencies below and overlapping with {\it Planck} frequency bands. Four samples, briefly described in  Sect.~\ref{sec:catalogue}, were selected from the Australia Telescope 20\,GHz (AT20G) survey catalogue (Murphy et al. 2010, Massardi et al. 2008, 2011a).
The PACO observations were carried out in several epochs between July 2009 and August 2010 in three pairs of 2\,GHz-wide bands (centered at 5.5 and 9\,GHz, 18 and 24\,GHz, 33 and 39\,GHz). At least one observation was made within 10 days from the {\it Planck} satellite observations in any of the LFI channels (30, 44, 70\,GHz).

In the present paper, after summarizing the selection criteria and the observation and data reduction procedures (\S\,\ref{sec:PACO}), we present the whole catalogue of observations in total intensity for all the 464 targets for which valid data were obtained in at least one PACO observing epoch
(\S\,\ref{sec:catalogue}). In \S\,\ref{sec:PCCS_fluxes} we describe the SEDs obtained combining PACO with quasi-simultaneous {\it Planck} data and we discuss the characterization of the radio source populations and their spectral properties over the 4.5-857\,GHz frequency range.
We finally summarize our findings in \S\,\ref{sec:conclusions}.

\section{The PACO project}\label{sec:PACO}
\begin{figure}
    \hspace{-2cm}
  \includegraphics[trim=1cm 2.5cm 0cm 2cm, clip,width=9.7cm, height=10cm, angle=90]{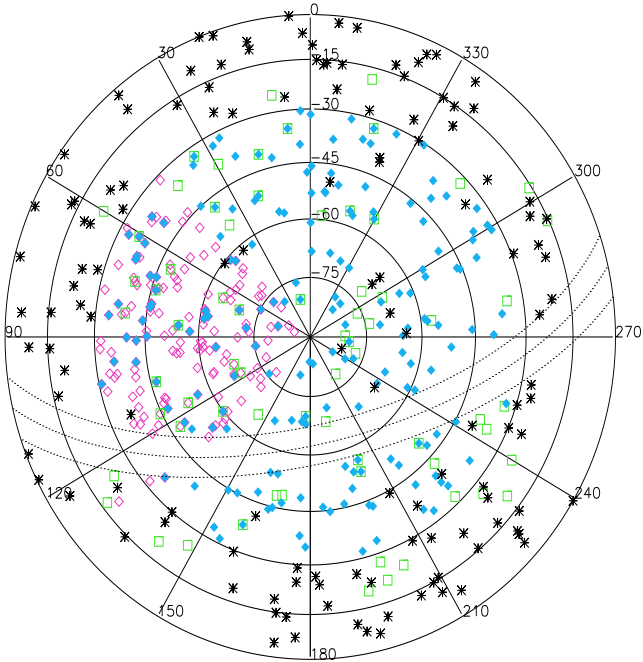}
  \caption{Polar projection of the Southern sky, showing the distribution of the PACO samples (RA=0\,h is at the top and the RA increases counterclockwise). The dotted lines indicate the Galactic plane and bound the area with Galactic latitude $|b|<5^\circ$. Sources in these regions were excluded from the PACO samples. The faint PACO sample (open pink diamonds) occupies the area with $3\hbox{h}<{\rm RA}<9$h. The bright PACO sample (filled blue diamonds) covers the whole area at $\delta<-30^\circ$. The spectrally-selected sample (green squares), the blazars and the AT calibrator sources (black asterisks) are spread over the whole southern sky.} \label{fig:map}
\end{figure}

The PACO samples were selected from the AT20G catalogue (Murphy et al. 2010) down to flux densities well below the expected detection limits for the first two {\it Planck} surveys in order to investigate both sources that could be detected by {\it Planck} and the contribution to confusion below the satellite detection limits. Poisson fluctuations due to unresolved extragalactic radio sources are, in fact, the main contaminant of {\it Planck} CMB maps on angular scales smaller than $20'$--$30'$ at frequencies of up to $\simeq 100\,$GHz and are dominated by sources just below the detection limit (Toffolatti et al. 1998, 1999, 2013; De Zotti et al. 1999). It is therefore important to characterize as accurately as possible such sources.

The final list of sources observed in the framework of the PACO project has 482 AT20G sources at Galactic latitude $|b|>5^\circ$ classified as extragalactic. As pointed out by Massardi et al. (2011b) the sample may still contain some Galactic sources but they should be very few because the relatively high resolution 20\,GHz observations preferentially select compact objects ($<30$\,arcsec) and Galactic sources with $|b|>1.5^\circ$ are rarely that small, especially at the brighter flux density levels. We have also excluded sources within a $5^\circ$ radius circle centered at $\hbox{RA}=80.8939^\circ$, $\delta=-69.7561^\circ$ to  avoid the Large Magellanic Cloud region where {\it Planck} flux densities are badly affected by confusion.

A comprehensive description of the observing strategy and of the data reduction procedures is in Massardi et al. (2011b). Here we present only a short overview.

The project got $\sim$450 hours allocated at the Australia Telescope Compact Array (ATCA) over 65 epochs between July 2009 and August 2010. Observations at 2 epochs have been discarded because of bad weather. Bad weather also seriously affected some of the Austral summer epochs making it impossible to get the 7 mm data in a few runs. A list of frequencies and configurations for each observing epoch is in Table 2 of Massardi et al. (2011b). All the 6$\times$22\,m antennas of the ATCA were used in all the runs. Time slots affected by bad weather or instrumental failures were flagged.

The Compact Array Broadband Backend (Wilson et al. 2011) system allows $2\times2$\,GHz simultaneous bands in continuum. We used the 7\,mm receivers with bands centered at 33 and 39\,GHz, to overlap the lower-frequency {\it Planck} channels, the 12\,mm receiver with bands centered at 18 and 24\,GHz, to include the AT20G selection frequency, and we extended the SEDs to lower frequencies using the 3-6\,cm receiver with bands at 5.5 and 9\,GHz. To better detail the source spectral behaviour we have split each 2\,GHz-wide band into $4\times 512$\,MHz sub-bands, and calibrated each sub-band independently.

The {\it Planck} scanning strategy has been publicly released soon after the satellite launch in May 2009. The satellite scans the sky in circles passing close to the ecliptic poles. All {\it Planck} receivers cross the position of sources lying on the ecliptic equator within a few days. Sources close to the ecliptic poles remain within the satellite focal plane for up to 2.5 months. The satellite covers the whole sky in about 6 months. Because the maps are the result of averaging all the scans through a position, we cannot get information from them on the very short term variability ($\lesssim 30$ days). Variability tests on the time ordered data are limited to the brightest objects because of lack of integration on source. We consider observations to be ``co-eval'' with the satellite if performed within 10 days from the satellite observations at any of its LFI frequencies. The {\it Planck} On-the-Flight Forecaster (Massardi \& Burigana 2010) has been developed and used to predict when the PACO sources are observed by the satellite, according to its pre-programmed pointing lists and the focal plane properties.

Data reduction was performed by a MIRIAD based pipeline (Sault et al. 1995) and on purpose developed code. For each epoch a suitable bright point-like bandpass calibrator and a primary calibrator (PKS1934-638 at any frequency or Uranus only at 7 mm) have been observed. Bandpass corrections have been applied to any target whose flux density was bootstrapped to the primary calibrator scale.

In previous work from our group (Massardi et al. 2011b) we identified a discrepancy in the flux density calibrations that uses Uranus or PKS1934-638 as primary calibrator arising from the models available for the sources at that epoch. In our data reduction we corrected the flux densities applying scaling factors that we estimated from observations of the two sources at our frequencies. Recently, Partridge et al. (2015) published a revised official version of the model for PKS1934-638 based on more precise high-frequency observations of Uranus, that corrects the calibration discrepancy. We corrected our calibrations applying new scaling factors according to the `new' model. Since the correction factor (with respect to the `old' model) is in the range 5.2\%-10.5\% for the frequencies we use in the 7mm band but less then 0.5\% at lower frequencies, we decided to apply it only to the 7mm band. This also allows a direct comparison in the 15 mm band with AT20G data, calibrated with the `old' model, for the variability purposes described below. 

Phases are self-calibrated as targets are expected to be pointlike and in the phase center.
Triple product techniques (Massardi et al. 2008) have been used to estimate the flux densities for compact sources. The amplitude of triple products is the geometric average of the visibility amplitudes in a baseline closure triangle and its phase is the phase closure: this method is particularly well suited for strong and point-like sources as it is able to recover the flux density lost in imaging because of phase decorrelation. Extended sources have been identified according to the classification in the AT20G catalogue and on the basis of the analysis of the phase closure at 7\,mm. For them the flux density is defined as the amplitude of the shortest spacings for the array configuration used. The noise term of the flux density errors for point sources was calculated as the root mean square (rms) of the triple product amplitudes over all the possible triangles of antennas. For extended sources the rms of the triple product has been multiplied by $\sqrt{n_{\rm b}}$ where $n_{\rm b}$ is the number of baselines of the used configuration.

A quality assessment procedure has been applied to the results of the automatic pipeline in order to identify further problems that might affect the data (unflagged bad weather conditions, some antenna correlations to be rejected, residual spikes in the band, etc.), and flag them a-posteriori. In the assumption that the true SEDs do not contain discontinuities, we compared the smoothness of the spectra of the data points with a fitting polynomial function with degree equal to the number of 2\,GHz bands observed. For each SED of the considered epoch we excluded those points (if any) that diverged from the fit more than 4 times the fractional rms divergence. After the quality assessment process, we were left with $\sim85$\,\% of our data.

The fractional rms divergence of the unflagged data for each epoch has been used as an estimate for the gain term of the noise. For the days with less than 10 good fits available we have used $\sigma_{\rm g}=0.012$, equal to the median of the rms fractional divergence calculated over all the epochs for point sources only, or $\sigma_{\rm g}=0.05$ for extended sources. The final error bars for each source in each epoch have been estimated by adding in quadrature the gain term fluctuations multiplied by the source flux density and the noise term.

\subsection{The PACO catalogue} \label{sec:catalogue}

The final PACO catalogue includes the 464 sources out of the 482 observed for which we have a valid flux density measurement (or a lower limit in case of extended objects) at least in one of the observing epochs. The new catalogue is available online as supplementary material. The 7 mm flux densities are estimated according to the new official model for PKS1934-638 (Partridge et al. 2015). Hence, the new catalogue updates also the data that we have already published for the 344 sources that form 3 partially overlapping complete sub-samples described in previous papers by our group:
\begin{itemize}
\item The ``bright PACO sample'' (Massardi et al. 2011b) comprises the 189 sources with AT20G flux densities $S_{20\rm{GHz}}>500\,$mJy at $\delta<-30^\circ$. The analysis of this sample demonstrated that only 14 per cent of the source SEDs could be described by a power law (74 per cent of which are flat spectra sources), while the large majority of the sample (66 per cent) is constituted of down-turning spectra sources, with median spectral index changing from -0.07 between 5 and 20\,GHz to -0.57 between 20 and 40\,GHz. Peaked sources constitute the 14 per cent of the sample, with a peak that is typically not very prominent and has a mean frequency equal to 16.3\,GHz. No source shows upturning or inverted spectrum over the whole frequency range. The sample shows a trend for an increasing variability with frequency and time lag.

\item The faint PACO sample (Bonavera et al. 2011), comprising the 159 sources with  AT20G flux densities $S_{\rm 20GHz}>200$\,mJy with right ascension $3<RA<9$\,hr and declination $\delta<-30^\circ$, showed that the fraction of steep spectrum sources increases from 3.6 per cent for sources with $S_{\rm 20GHz}>500$\,mJy to 13.3 per cent for sources with $S_{\rm 20GHz}<500$\,mJy. The comparison with the AT20G observations, performed on average 2--4 years earlier, indicates a high 20\,GHz variability with mean $V_{\rm rms}\simeq 20$ per cent (see Sect. \ref{sec:PACO_fluxes}). The large range of spectral behaviours in the range 5--40\,GHz and the high variability rate confirmed the importance of simultaneous observations to reconstruct radio source SEDs. The faint sample has been selected to be approximately 5 times fainter than the expected {\it Planck} detection limits \citep{Leach2008} at the lower frequencies  to investigate source confusion that is a serious issue for {\it Planck}, due to its rather large beam sizes. Sources near the {\it Planck} detection limit may be strongly affected by the Eddington's (1913) bias that leads to systematic flux density overestimates (Hogg \& Turner 1998). Also, confusion fluctuations may produce high intensity peaks that may be misinterpreted as real sources. Moreover, confusion can shift the positions of intensity peaks from the true source positions (Hogg 2001), complicating their identification. An accurate control of these effects requires simultaneous ground-based observations with much better resolution and signal-to-noise ratio. The faint sample has been successfully used to confirm the quality assessment for the {\it Planck} Early Release Compact Source Catalogue (ERCSC, Planck collaboration 2011) and to estimate the radio source contribution to CMB power spectra down to $\sim 200$\,mJy at 30 and 44\,GHz.
The faint PACO sources in a region surrounding the Ecliptic pole were recently followed-up in a polarimetric dedicated observation in all the PACO bands (Galluzzi et al. in preparation).

\item The ``spectrally-selected PACO sample'' (Bonaldi et al. 2013) comprises the 69 sources with  AT20G flux densities $S_{20\rm{GHz}}>200$\,mJy and spectra classified by Massardi et al. (2011b) as inverted or upturning in the frequency range 5--20\,GHz, selected over the whole Southern sky. We used the variability of sources in the spectrally selected sample between AT20G (2004--2007) and PACO (2009--2010) epochs to discriminate between candidate High-Frequency Peakers (HFPs) and candidate blazars. The HFPs selected by our criteria have spectral peaks $>10$\,GHz in the observer frame and turn out to be rare ($<0.5$ per cent of the $S_{20 \rm{GHz}}\ge 200$\,mJy sources), consistent with the short duration of this phase implied by the `youth' scenario. Most ($\simeq 89$ per cent) blazar candidates have remarkably smooth spectra, well described by a double power law, suggesting that the emission in the PACO frequency range is dominated by a single emitting region. Sources with peaked PACO spectra show a decrease of the peak frequency with time at a mean rate of $\sim3 \pm 2$\,GHz\,yr$^{-1}$ on an average time-scale of $\langle \tau \rangle = 2.1 \pm 0.5$\,yr (median: $\tau_{\rm median} = 1.3$\,yr). The 5--20\,GHz spectral indices show a systematic decrease from AT20G to PACO. At higher frequencies spectral indices steepen: the median spectral index between 30 and 40\,GHz is steeper than the one between 5 and 20\,GHz by $\delta \alpha = 0.6$. Taking further into account the Wide-field Infrared Survey Explorer data we found that the SEDs, $\nu S(\nu)$, of most of our blazars peak at $\nu_{\rm {SEDp}} < 10^5$\,GHz; the median peak wavelength is $\lambda_{\rm{SEDp}} \simeq 93\,\mu$m. Only six have $\nu_{\rm{SEDp}} > 10^5$\,GHz.
\end{itemize}

Moreover the PACO sample comprises 203 ATCA calibrators (76 of which are included in the above mentioned complete samples) that showed more than 10\,\% variability at 20\,GHz in the epoch 2006--2009 and with 20\,GHz flux density $S_{20\rm{GHz}}>200$\,mJy. Most of them were part of long-term monitoring programs with several instruments including APEX, Metsahovi, ATCA and Effelsberg. By including these sources into the PACO project we aimed at rising the probability to get simultaneous observations over a wide frequency range of flaring events detected by {\it Planck}.

Nineteen sources are extended and, even if included in the catalogue with a proper flag, are excluded from the following analysis because the PACO flux densities could be underestimated as the triple-product technique is better suited for point-like sources. The final sample included in the following analysis has 445 sources.

\begin{table}
\caption{Mean variability index as a function of time lag and frequency for the PACO  brigh sample (BS), faint sample (FS) and spectrally-selected sample (SS). The time lag of 2--4\,yr refers to the comparison between PACO and AT20G.}\label{tab:var_index}
\begin{tabular}{cccccccc}
\hline
Sample   & Time lag & &&Band&[GHz]&&\\
&& 5& 9&18&24&33&39\\
\hline
  FS     &    90 days  &5.9 &7.4 &7.3 &7.8 &8.5&8.9              \\
  FS     &    180 days &6.7 &9.3 &10.6&10.1&14.9&5.7              \\
  FS     &    270 days &5.2 &7.4 &7.3 &7.9 &11.7&13.5             \\
  FS     &    360 days &9.8 &8.9 &8.2 &11.8&12.0&13.3            \\
  FS     &    2--4 yr  &24.9&25.0&27.2&& &               \\
\hline
  BS     &    90 days  &4.8 &6.2 &7.4 &6.9&9.5&9.2  \\
  BS     &    180 days &5.7 &7.0 &8.9 &8.1&6.8&6.2  \\
  BS     &    270 days &6.2 &6.9 &7.6 &7.4&12.4&14.2 \\
  BS     &    360 days &7.5 &11.8&10.5&12.0&10.7&11.0            \\
  BS     &    2--4 yr  &18.1&19.4&19.9&& &               \\
\hline
  SS     &    90 days  &6.2 &7.0&6.9&9.4&7.1  &7.6              \\
  SS     &    180 days &7.9 &7.7&8.9&10.2&8.5&10.8              \\
  SS     &    270 days &6.1 &7.9&11.2&13.1&14.6&15.9   \\
  SS     &    360 days &6.9 &8.3&11.9&14.0&16.3&16.1  \\
  SS     &    2--4 yr &15.4&17.3&23.2&& &               \\
\hline\end{tabular}
\end{table}
\section{Combining PACO and PCCS data} \label{sec:multifreq}

\subsection{PACO flux densities} \label{sec:PACO_fluxes}
As already mentioned, the PCCS flux densities are the result of repeated observations of the same source within the {\it Planck} nominal mission (from July 2009 to November 2010). The number of observations per source is a function of the source position, with a peak close to the Ecliptic pole, where sources are observed continuously for about three months for each survey, and a minimum at the equator, where observations last for few days every 6 months (see Fig. 5 in Massardi \& Burigana 2010). Comparing {\it Planck} and PACO observations is therefore not trivial, and source variability could still be an issue despite the simultaneity of the PACO observations with {\it Planck} scans.

For highly variable objects the effect of PCCS flux density averaging is twofold: on the one hand it emphasizes the presence of long lasting flaring events, and on the other it can average out short term variability. In both cases, to minimize the effects of variability, the observations should be ideally performed and averaged over the same epochs. Even in the case of sources on the Ecliptic equator, the whole {\it Planck} field of view takes no less than 7 days to move past its position on the sky. On these timescales we do not expect source variability to be larger than few percent at any frequency.

Information on the typical variability of the sources of our samples can be obtained by comparing the AT20G--PACO flux densities and the PACO flux densities for different epochs.
Following Sadler et al. (2006) we define the variability index as:
$$
V_{\rm rms}=\frac{100}{\langle S \rangle}\sqrt{\frac{\sum [S_i-\langle S \rangle]^2-\sum\sigma^2_i}{n}},
$$
where $S_i$ is the flux density of a given source measured at the i-th epoch, $\sigma_i$ is the associated error, $n$ is the number of measurements, and $\langle S \rangle$ is the mean flux density of the source, computed using all the available observations (not only those used for computing the variability index).  In Table~\ref{tab:var_index} we report the mean variability indices of our samples for different time lags and frequencies.
%
%

In previous papers from our collaboration based on the PACO data only (Massardi et al. 2011b; Bonavera et al. 2012) we confirmed a trend toward an increase of the variability amplitude with frequency (Impey \& Neugebauer 1988; Ciaramella et al. 2004) and a marginal indication of higher variability for the longer time lag. At $\sim 20$\,GHz the median variability index for sources with $S_{\rm 20 GHz}>500$\,mJy for a 6 months lag is $\simeq 6\,\%$, and grows to $\simeq 7\,\%$ at $\simeq 35$\,GHz. The values become $\simeq 9\,\%$ and $\simeq 11\,\%$ respectively for 9 months time lag. The comparison between AT20G and PACO observations gave a weak indication that variability increases with flux density, at least over time lags of a few years.

Given the non-negligible variability of the sources within the {\it Planck} nominal mission period, the best PACO flux densities to be compared with PCCS flux densities are a suitable average of the PACO data for different epochs. There are 34 sources having only data for one PACO observing epoch.
We excluded them from the generation of multifrequency fits to improve the homogeneity between {\it Planck} and PACO data.

We first averaged the flux densities of the $4\times 512$\,MHz sub-bands, obtaining for every PACO epoch 6 values corresponding to the $6\times2$\,GHz observing frequency bands. These larger bandwidths are more directly comparable with the large {\it Planck} spectral bands. We defined the error for each mean flux density as the maximum measured error for the 4 sub-bands.
We then obtained a single flux density for each source at each of the six PACO spectral bands by averaging over all the PACO observing epochs available, with an error given by the average error of the individual observations.
For each frequency band we also calculated a contribution to the error due to the source variability, defined as the standard deviation of the flux densities corresponding to the different epochs. The use of this error component is explained in the following sections.


\begin{table}
\caption{Number of detection of pointlike sources listed in PACO catalogue in the PCCS for each photometric method and {\it Planck} frequency. For each method we also considered the detections among the PACO Bright and Faint Sample (i.e. for the complete samples selected with $S_{\rm 20\,GHz}>0.5$ or $S_{\rm 20\,GHz}>0.2 $\,Jy respectively).}\label{tab:planck_det}
\begin{tabular}{lccccccc}
\hline
Method    & Tot & & & Band&[GHz]&\\
Sample   &  & 30& 44&70&100&143&217\\
\hline
DETFLUX&&&&&&\\
 all $>5\sigma$&445 &224  &   104  &   141   &  226   &  238   &  223   \\
 all $>3\sigma$&445 &259  &   149  &   175   &  282   &  311   &  282   \\
 BS $>3\sigma$&174 & 136   &    79    &  92 &     143  &    152 &     141  \\
 FS $>3\sigma$&143 &46      & 30     &  36    &   67   &    78    &   70    \\
APERFLUX &&&&&&\\
 all $>5\sigma$&445 &68  &   22  &   91   &  121   &  155   &  192   \\
all $>3\sigma$&445&255 &     147   &   175   &   281   &   309  &    281   \\
BS $>3\sigma$&174&134    &   78    &  92    &  143    &  150 &     141   \\
FS $>3\sigma$&143&45    &   29    &   36     &  66    &   77   &    69    \\
PSFFLUX&&&&&&\\
 all $>5\sigma$&445 &249  &   140  &   150   &  220   &  230   &  218   \\
all $>3\sigma$&445&255  &    149  &    175   &   281 &     311  &    280  \\
BS $>3\sigma$&174&136    &   79   &   92    &  143  &    152  &    141  \\
FS $>3\sigma$&143&44     &  30    &   36     &  66 &      78    &   69  \\
GAUFLUX&&&&&&\\
 all $>5\sigma$&445 &253  &   112  &   138   &  211   &  219   &  174   \\
 all $>3\sigma$&445&  258   &    117    &   159    &   265    &   287   &    216  \\
 BS $>3\sigma$&174&136   &     57    &    81   &    133    &   138    &   107  \\
FS $>3\sigma$&143&45    &    18    &    33    &    64     &   72    &    52   \\
\hline\end{tabular}
\end{table}

\subsection{PCCS flux densities: comparison of photometric methods} \label{sec:PCCS_fluxes}

The PCCS flux densities are the result of source extraction from the {\it Planck} full sky maps of the nominal mission. Several extraction and photometric methods were applied for cross-check and validation purposes. More details about each method are given in Planck Collaboration XXVIII (2014). Thanks to the simultaneity of PACO and {\it Planck} observations, we can exploit our data to perform an additional assessment of the performances of the photometric methods, at least at the frequencies that overlap with our bands.

We used the 1.3 release of the PCCS for the LFI data and the 1.2 release for the HFI data. We cross-matched each source in our catalogue with the PCCS catalogues with a search radius equal to half the {\it Planck} FWHM for each {\it Planck} frequency. We did not find any multiple association.

The PCCS gives, for each source, four different measures of the flux density: Detection pipeline photometry (DETFLUX), Aperture photometry (APERFLUX), Point Spread Function (PSF) fit photometry (PSFFLUX) and Gaussian fit photometry (GAUFLUX).

\begin{figure*}
\begin{center}
\includegraphics[trim=0cm 0cm 10cm 0cm, clip,  width=6cm, angle=90]{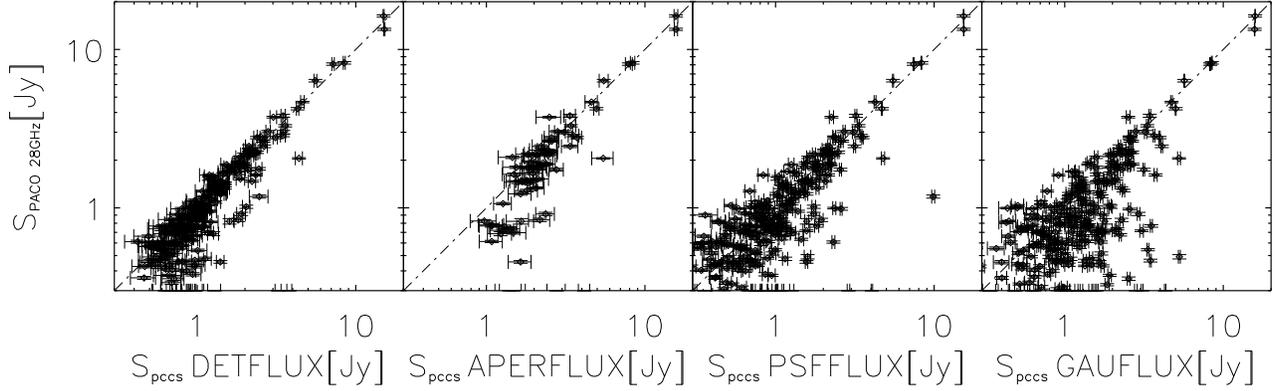}
\caption{Comparison of the PACO integrated flux densities at $\sim30$\,GHz (obtained averaging PACO 33 and 24\,GHz flux densities) and PCCS flux densities obtained with the detection pipeline photometry, the aperture photometry, the PSF photometry, and the Gaussian fit photometry (respectively from the left to the right).}\label{fig:paco2pccs_30}
\end{center}
\end{figure*}

\begin{figure*}
\begin{center}
\includegraphics[trim=0cm 0cm 10cm 0cm, clip,  width=6cm, angle=90]{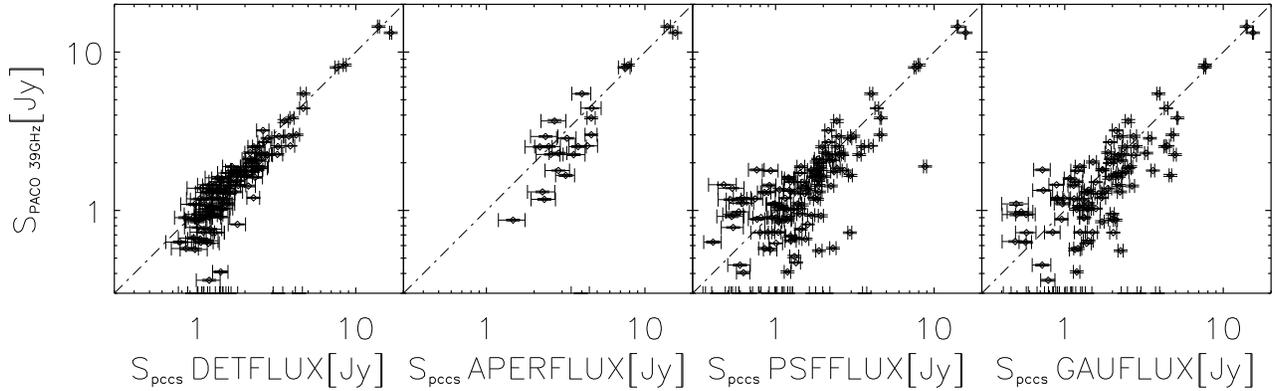}
\caption{Comparison of 39\,GHz PACO integrated flux densities and 44\,GHz PCCS flux densities obtained with the detection pipeline photometry, the aperture photometry, the PSF photometry, and the Gaussian fit photometry (respectively from the left to the right).}\label{fig:paco2pccs_44}
\end{center}
\end{figure*}

Table~\ref{tab:planck_det} lists the number of detections above 3\,and 5\,$\sigma$ for the different PCCS photometric methods at {\it Planck} frequencies up to 217\,GHz. There is no significant difference in the fraction of detections among the different methods at a low significance level, while some differences arise in the number of detections at higher signal-to-noise ratios.

In Figs.~\ref{fig:paco2pccs_30} and \ref{fig:paco2pccs_44} we show the comparison between PACO flux densities and PCCS flux densities at $\sim 30$ and 44\,GHz for the first 4 PCCS photometric measures. Since the central PCCS frequency in the $\sim30$\,GHz band is at 28\,GHz we averaged the PACO 33 and 24\,GHz flux densities for comparison purposes. At 44\,GHz we compared the 39\,GHz band of PACO with the PCCS data centered at 44.1\,GHz without any correction for the slightly different frequency.

Table \ref{tab:planck_cfr} shows, for each photometric method, the mean relative difference between PACO and PCCS flux densities for bright sources ($>1$ Jy in both the catalogues) detected above the 5$\sigma$ significance level in PACO.

There is an overall good agreement between PACO flux density measurements and the PCCS DETFLUX photometry at 30\,GHz and at 44\,GHz. At both frequencies the {\it Planck} flux densities below the 90\% completeness levels quoted by \citet{PCCS} are boosted by the Eddington bias:  faint sources are preferentially detected if they lie on top of positive background (noise plus confusion) fluctuations. Hence, on average, their flux densities are over-estimated. We also note that the results of the present comparison of PCCS/DETFLUX with ATCA flux densities are,  for both frequencies, consistent with those found by Partridge et al. (2015) for smaller samples of sources observed with the VLA and the ATCA but using \textit{Planck} flux densities derived from purpose-made maps including only data taken during the period of ground based observations.
This indicates that our approach efficiently minimizes the effects of variability allowing us to perform the multifrequency analysis presented in the following.

The other PCCS photometric measures are either affected by larger error bars, implying a relatively low number of $5\,\sigma$ detections (APERFLUX) or show larger rms differences with PACO measurements (PSFFLUX). GAUFLUX is appropriate for extended sources \citep{PCCS} while our sources are point-like at the \textit{Planck} resolution. Hence in the following analysis we will adopt the DETFLUX photometry.

We have checked the 16 point-like sources (J010645-403419, J030956-605839, J033930-014635, J040353-360500, J045314-280737, J045703-232451, J050644-610941, J062923-195919, J065024-163739, J114701-381210, J142741-330531, J145726-353910, J162546-252739, J162606-295126, J183339-210341, J191109-200655) that showed PCCS to PACO flux density ratios larger than 1.5 at either 30 or  44\,GHz.  By inspecting the {\it Planck} maps we have found that in 11 cases there is a clear evidence of extended emission in the {\it Planck} maps in the immediate vicinity of the source. This emission, which can be attributed to the Galaxy, has likely contaminated the PCCS photometry.

\begin{table}
\caption{Mean and standard deviation of the relative difference (in per cent) between PACO and PCCS flux densities at $\sim 30$ and $\sim 40$ GHz for each {\it Planck} photometric method. Only {\it Planck} detections above $5\sigma$ and, to avoid Eddington bias contamination, with flux densities above 1 Jy both in PACO and PCCS have been considered .}\label{tab:planck_cfr}
\begin{tabular}{lcccc}
\hline
Method    & $ \sim30$ GHz & $\#$ & $\sim40$ GHz & $\#$\\
\hline
DETFLUX  & $0.8\pm 1.6$ & 89 & $7.9\pm 1.7$ & 83  \\
APERFLUX & $6.7\pm 2.6$ & 53 & $12.1\pm 5.6$ & 22  \\
PSFFLUX  & $1.1\pm 2.5$ & 83 & $5.3\pm 2.7$ & 70  \\
GAUFLUX  & $6.6\pm 2.5$ & 84 & $7.1\pm 3.5$ & 63  \\
\hline\end{tabular}
\end{table}

\begin{figure*}
  \includegraphics[width=17cm]{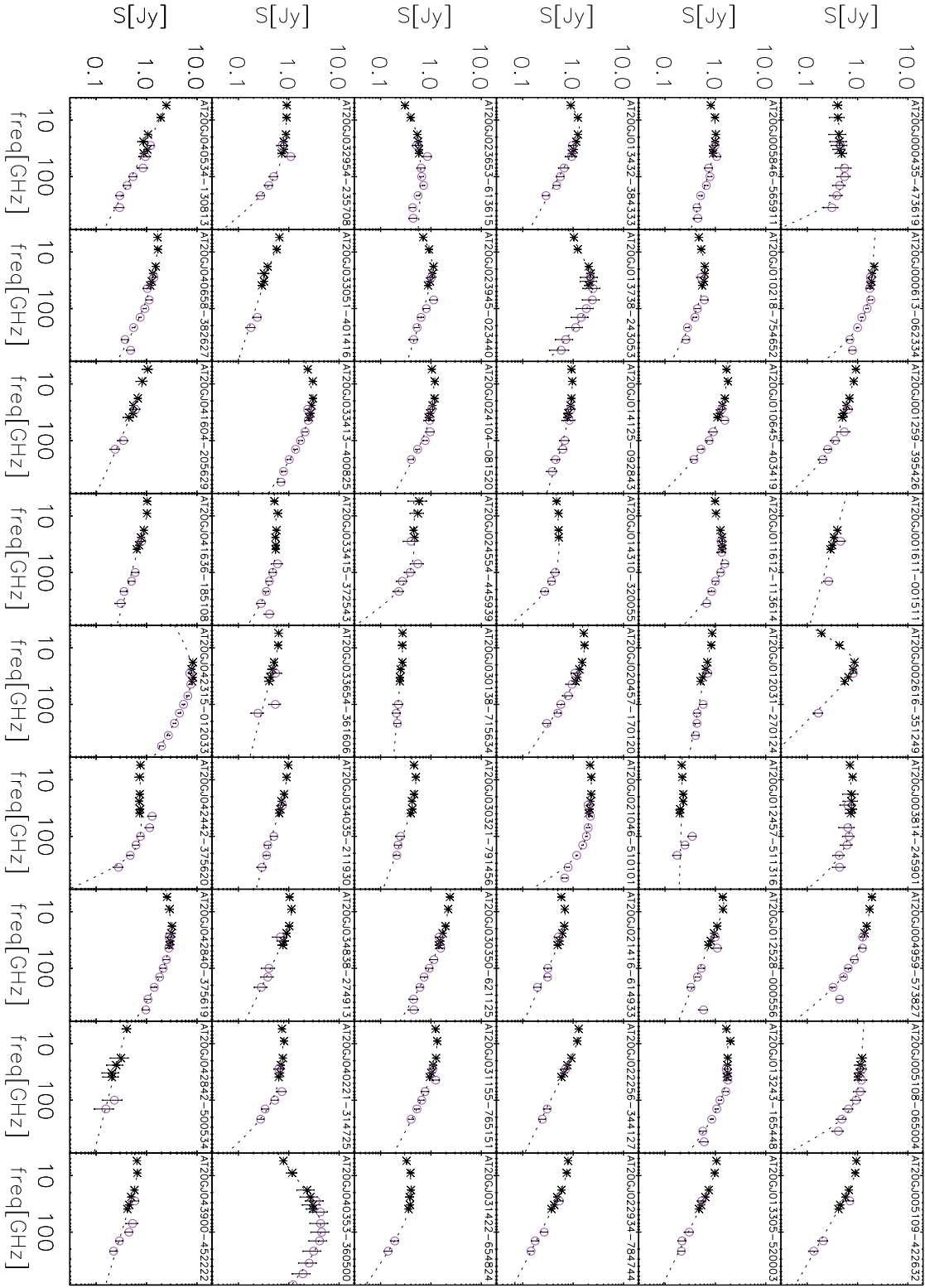}
 \caption{Averaged PACO spectra of sources with at least 2 PACO observations and 2 points in the {\it Planck} frequency bands (open circles) up to 217\,GHz fitted with a double power law (dotted line).}\label{fig:SED}
\end{figure*}
\addtocounter{figure}{-1}
\begin{figure*}
  \includegraphics[width=17cm]{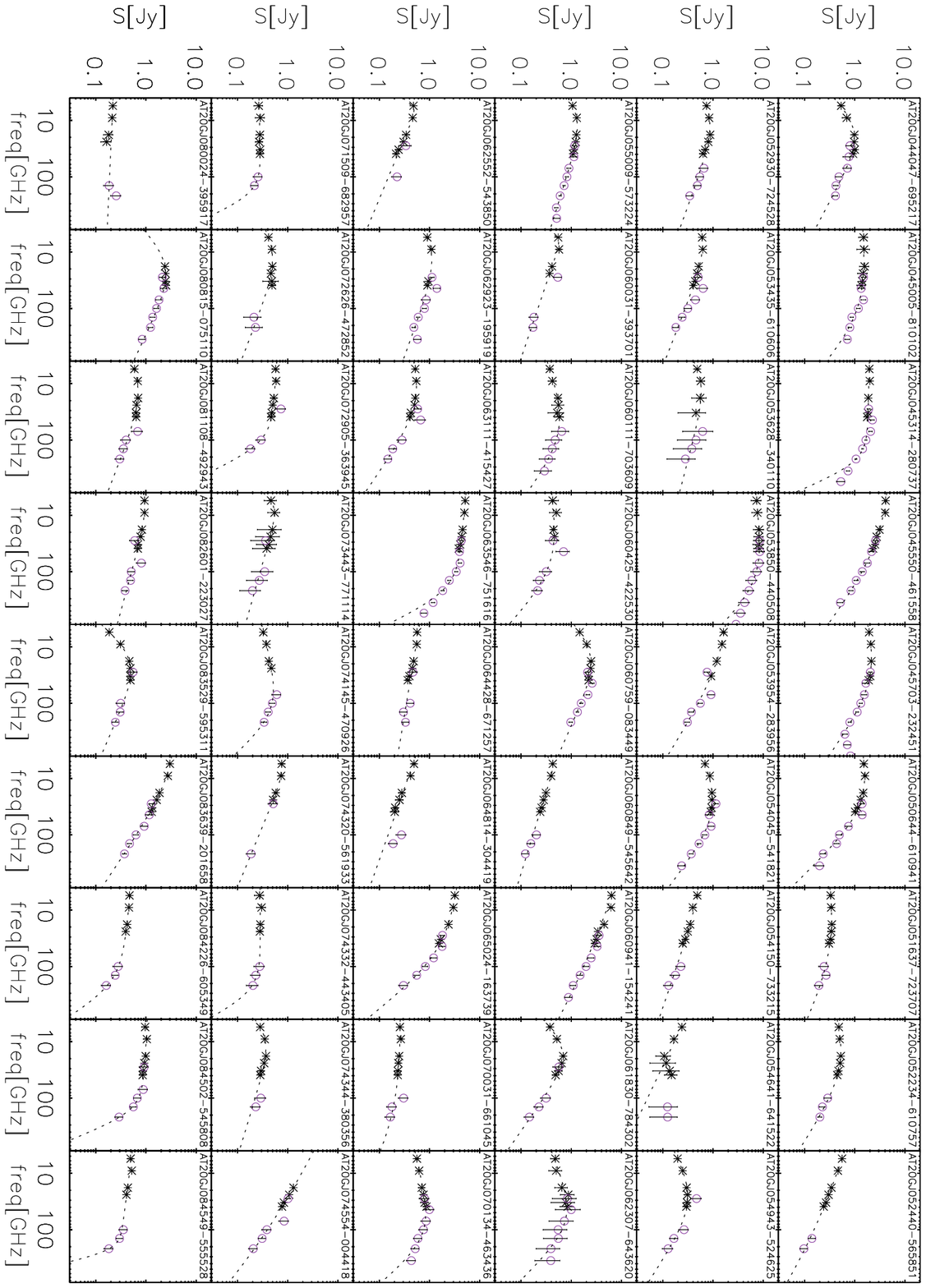}
  \caption{Continued.}
\end{figure*}
\addtocounter{figure}{-1}
\begin{figure*}
  \includegraphics[width=17cm]{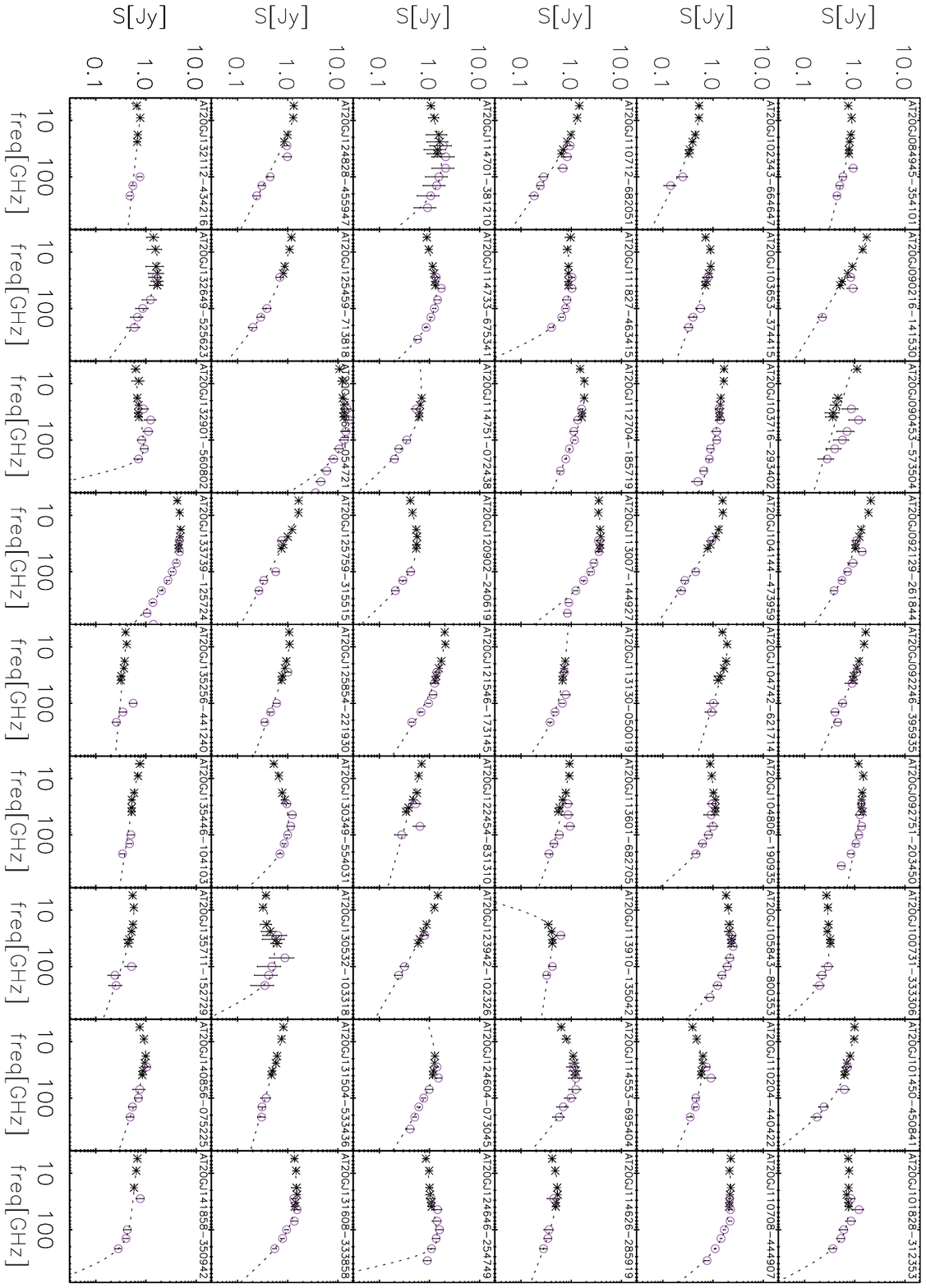}
  \caption{Continued.}
\end{figure*}
\addtocounter{figure}{-1}
\begin{figure*}
  \includegraphics[width=17cm]{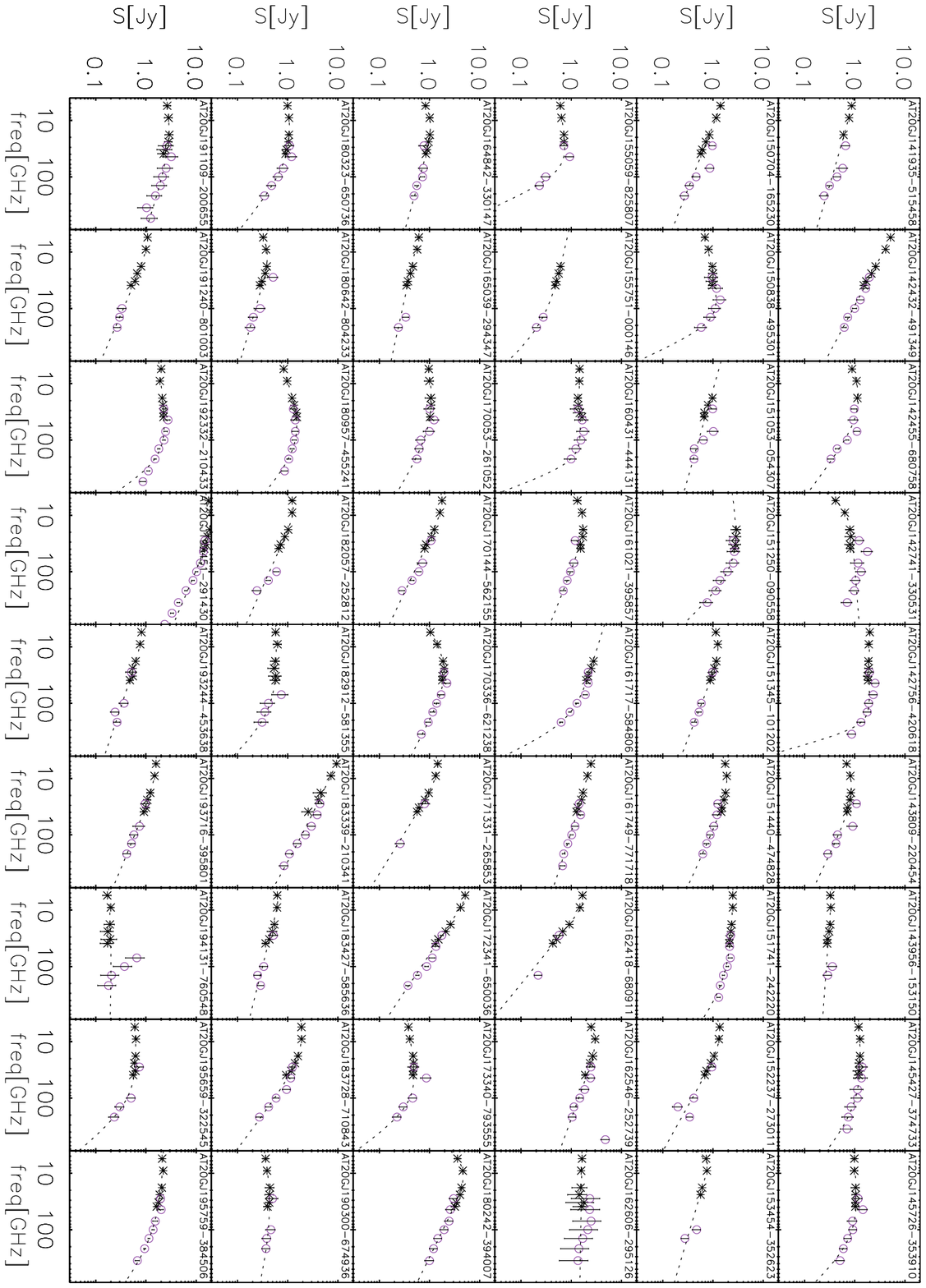}
  \caption{Continued.}
\end{figure*}
\addtocounter{figure}{-1}
\begin{figure*}
  \includegraphics[width=17cm]{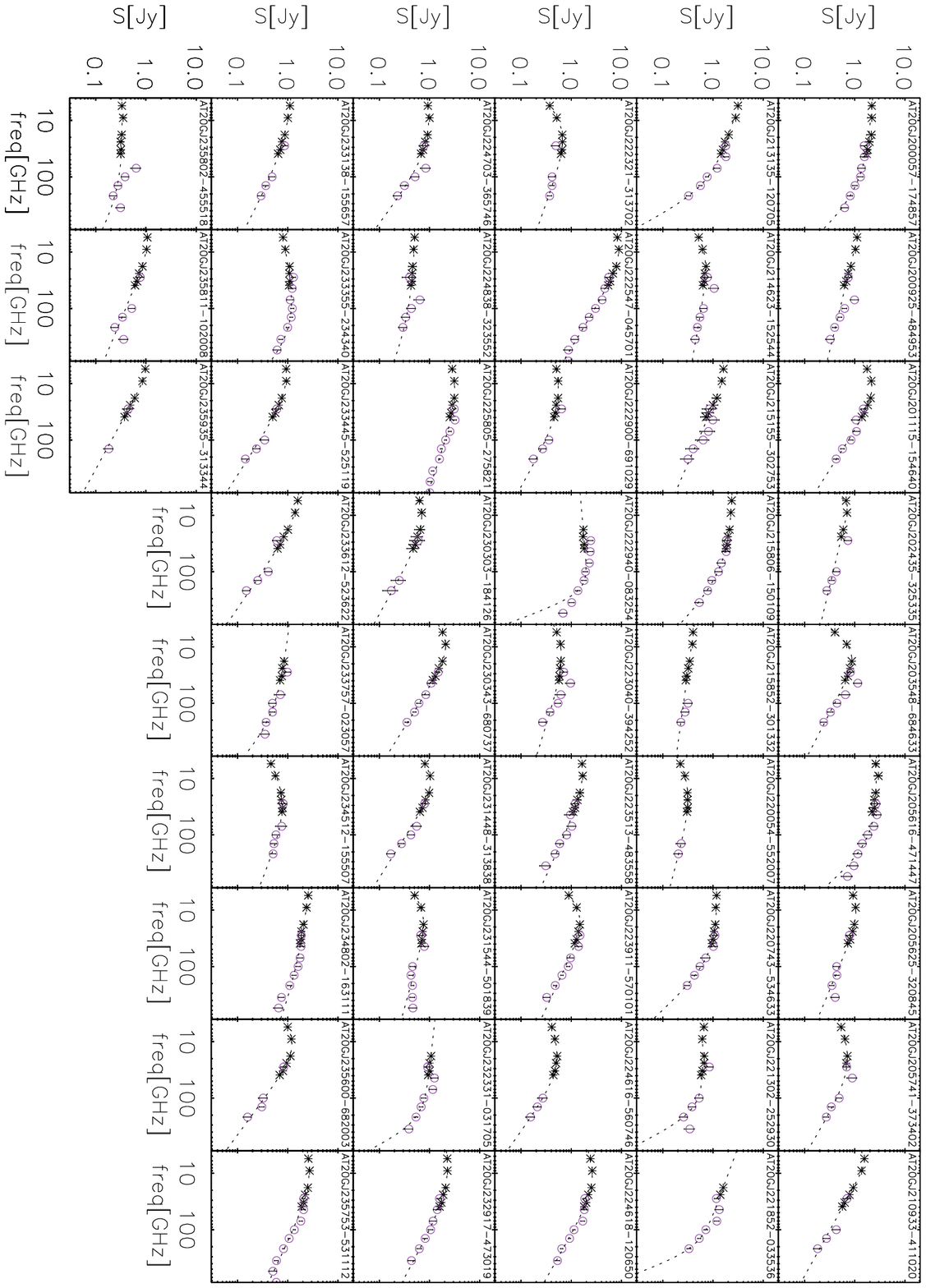}
  \caption{Continued.}
\end{figure*}

\subsection{Multifrequency spectra} \label{sec:fit}
In previous papers from our group (Massardi et al. 2011b; Bonavera et al. 2012; Bonaldi et al. 2013) we found that the majority of our spectra are well described in the 5--40\,GHz range by a double power-law. In this section we want to verify whether this is still a good description of the spectra over a broader range that includes both PACO and {\it Planck} data. Consistently with our previous work, the relation we fitted to the data is

\begin{equation}
S(\nu)=S_0/[(\nu/\nu_0)^{-a}+(\nu/\nu_0)^{-b}]
\end{equation}
where $S_0$, $\nu_0$, $a$ and $b$ are free parameters. The fit has been performed in logarithmic units by minimizing the $\chi^2$ with a nonlinear optimization technique based on an implementation of the Generalized Reduced Gradient optimization method.

We fitted this relation to the PCCS/DETFLUX data and to the averaged PACO spectrum obtained as described in Sect.~\ref{sec:PACO_fluxes}. We considered in the fit only the PCCS flux densities at frequencies up to and including 217\,GHz. In fact, at higher frequencies there are hints that the measured flux densities of several sources may be contaminated by Galactic thermal dust emission.
There are 290 sources detected at least in  two {\it Planck} bands up to 217\,GHz and for which we could define the averaged PACO spectrum; 147 of them are in the bright PACO sample, 31 in the spectrally-selected sample and 68 in the faint sample.

The errors on the averaged PACO flux densities were estimated summing in quadrature the mean errors of individual measurements and the standard deviation of the measurements at the different epochs, taken as estimates of the effect of source variability.  Similarly, we assigned to the PCCS flux densities error bars given by the sum in quadrature of nominal errors with the contribution due to variability estimated by PACO at 39\,GHz. This is likely a lower limit to the variability contribution at higher {\it Planck} frequencies, given that variability is expected to increase with frequency.

The mean reduced $\chi^2$ of the fits is 0.87 with a standard deviation of 0.55. This means that the double power-law model is still a good representation for most of the spectra over the wider frequency range. The exception are 26 sources ($\simeq\,9\%$) for which $\chi^2$ is above the value corresponding to 20\% probability threshold for 6 degrees of freedom as on average for our sources. These have been excluded from the following statistical analysis.

The composite spectra and the fits are shown in Fig.~\ref{fig:SED} for the remaining 264 sources.

\subsection{Global spectral behaviour}
\begin{figure}
\begin{center}
\includegraphics[width=8cm, angle=90]{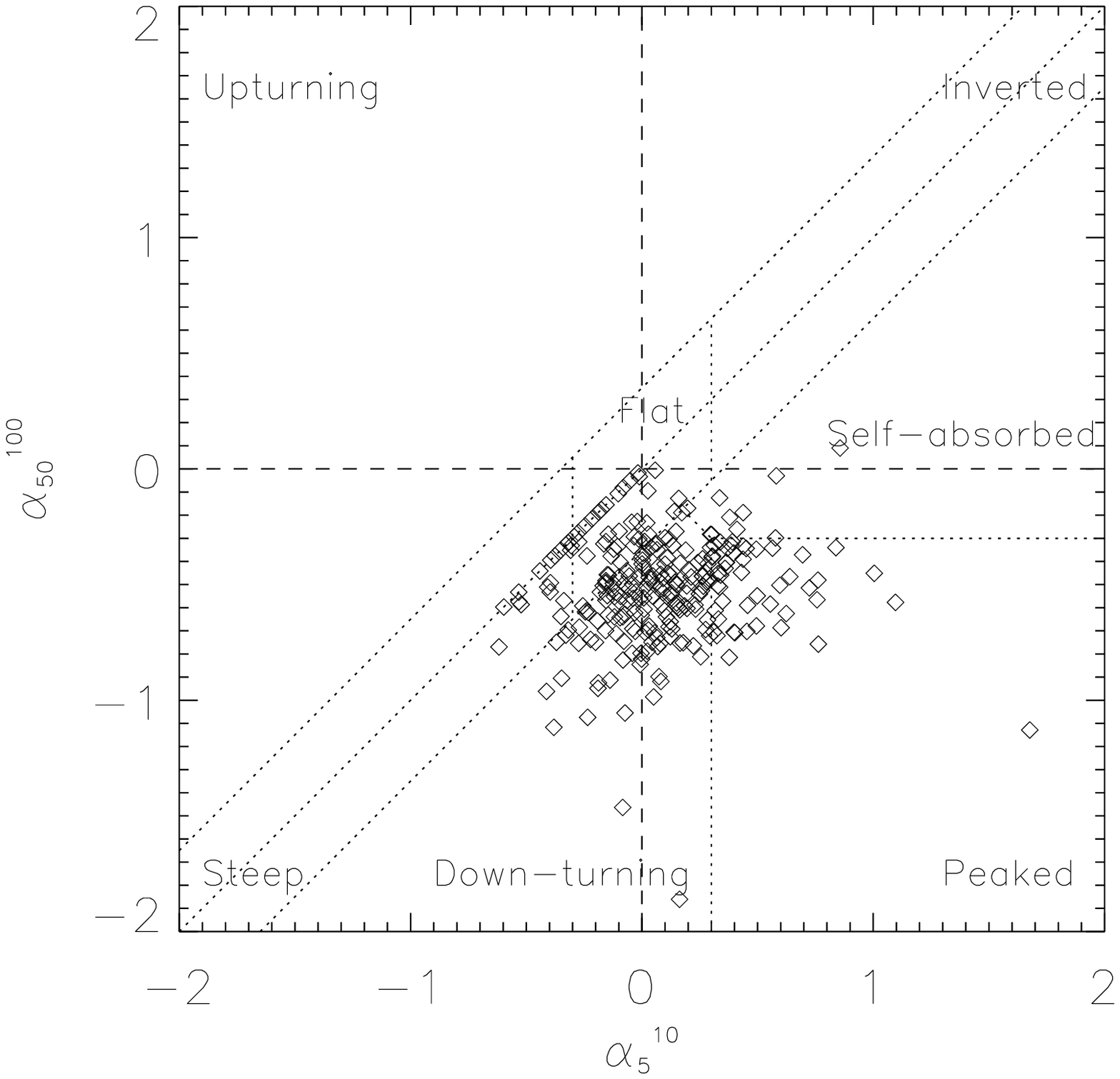}
\caption{Colour-colour plot comparing the spectral indices of the fitting double power-law function over the 5-10 and 100-200 GHz ranges.\label{fig:alpha_beta}}
\end{center}
\end{figure}

Figure~\ref{fig:alpha_beta} compares the slopes of the spectral indices of the fitting double power-law function over the 5--10 and 50--100\,GHz ranges.

The overall spectra can be classified into two categories: peaked (self-absorbed) and down-turning. Sources in the first category are defined as having $\alpha_5^{10}>0.3$ and $\alpha_{100}^{200}<-0.3$; there are 47 such sources. 

Mahony et al. (2011) report redshifts for 291 sources in the PACO catalogue; for 199 of them we have obtained a fit of the spectrum including PCCS data. The observed  median peak frequency for the peaked sources is 18.3 GHz with first and third quartile respectively equal to 12.8 and 27.3 GHz. For the 29 peaked sources with a redshift determination the median rest frame peak frequency is 39.6 with first and third quartile respectively equal to 31.4 and 53.0.

\begin{figure}
\begin{center}
\includegraphics[width=8cm, angle=90]{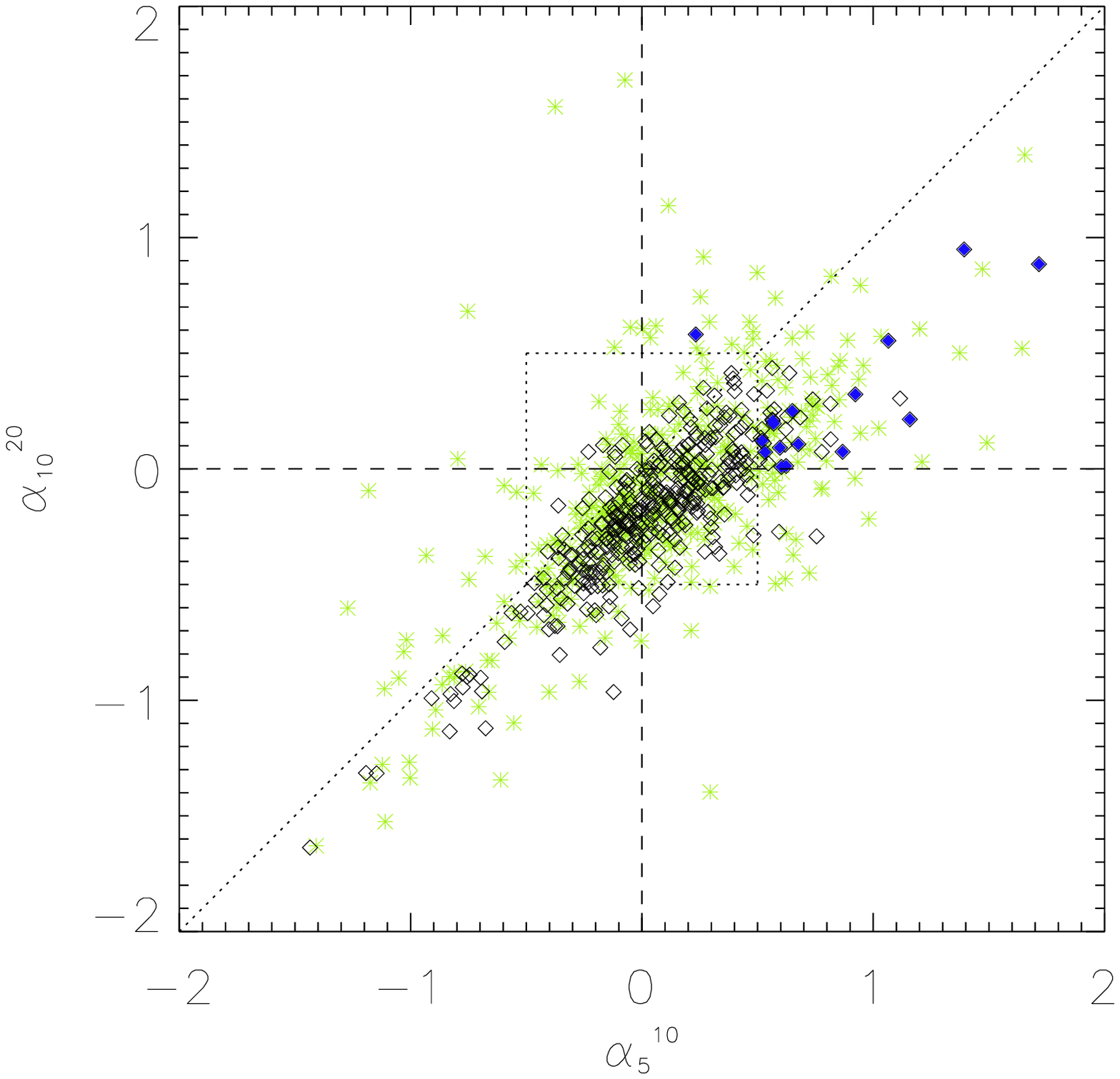}
\caption{Colour-colour plot comparing the $\sim5$--10\,GHz and $\sim10$--20\,GHz spectral indices for the PACO best epoch data (black diamonds) and the AT20G data (green asterisks). Filled blue diamonds represent the objects that remained in the inverted-self-absorbed region of the plot at both observing epochs and hence are candidate to be GPS sources.\label{fig:cc_paco_at20g}}
\end{center}
\end{figure}

The colour-colour plot comparing the $\sim5$--10\,GHz and $\sim10$--20\,GHz spectral indices could be used to identify candidate HFP sources as those maintaining inverted or self-absorbed spectra over a long period, i.e. showing low variability (see Fig.~\ref{fig:cc_paco_at20g}). The comparison of the spectral indices between the AT20G catalogue and the PACO best epoch of observation identifies a sample of 16 candidate high frequency peaking objects with this property. After removing the 6 sources identified as candidate blazar-like radio sources in WISE (D'Abrusco et al. 2014), in Massaro et al. (2014), or Healey et al. (2008) and the candidate high frequency peaking source (AT20GJ002616-351249) classified as a beamed source in Hancock et al. (2010), 9 sources remain as genuine candidate high frequency peaking objects: AT20GJ025055-361635, AT20GJ042810-643823,  AT20GJ045316-394905,  AT20GJ081240-475153, AT20GJ104742-621714, AT20GJ114553-695404, AT20GJ124345-681103, AT20GJ145508-315832, and AT20GJ183923-345348.


A visual inspection of Fig.~\ref{fig:SED} highlights remarkable properties of the spectra. Most sources are classified as flat-spectrum at cm wavelengths. A widely held view is that their spectra result from a superposition of a number of compact regions of different sizes, self-absorbed at different frequencies (e.g. Marscher et al. 1980; see De Zotti et al. 2010 for a review). However, with few exceptions (26 out of the 290 sources), the spectra are remarkably smooth over about two decades in frequency, suggesting that the radio to mm emission of most sources comes from a single component. We note, in this respect, that our samples of flat-spectrum sources are dominated by blazars, generally interpreted as compact radio source whose relativistic jets are observed along lines of sights making small angles, $\theta$, to the jet velocity, $\vec{V}$. According to the Blandford \& K{\"o}nigl (1979) model, if these sources are made of different parts moving with different speeds, a distant observer will see predominantly those parts that are moving with $\beta=V/c\sim \cos(\theta)$. This is consistent with VLBA \citep[e.g.,][]{Lister2009} and VLBI \citep[e.g.,][]{Lu2012} observations showing that, at a few cm to mm wavelengths, blazars flux densities are frequently dominated by a compact component with a smooth intensity profile.

The vast majority of spectra are steep above 30\,GHz, and can thus be interpreted as due to optically thin synchrotron emission. In other words, the canonical categorisation of radio sources as flat- or steep-spectrum breaks down at mm wavelengths. This means that spectral properties can no longer be used to separate the source population into compact and extended sources. A hint of a steepening above 30\,GHz in the rest frame was reported by Chhetri et al. (2012) using data at frequencies $\le 20\,$GHz.

Another noteworthy property is the lack of indication of a spectral break at still higher frequencies. Such break is expected as the effect of `electron ageing' (Kardashev 1962, Murgia et al. 2002). If the radio source is continuously replenished by a constant flow of fresh relativistic particles with a power law energy spectrum, $N(E)\propto E^{-\delta}$ (continuous injection model) the radio spectrum has a spectral index $\alpha_{\rm inj} = -(\delta -1)/2$ below a critical frequency $\nu_{\rm br}$ and $\alpha_{\rm h} = \alpha_{\rm inj} - 0.5$ above $\nu_{\rm br}$. If there is no expansion and the magnetic field is constant, the frequency $\nu_{\rm br}$ (in GHz) depends on the source age, $\tau_{\rm syn}$ (in Myrs), on the intensity of the magnetic field $B$ (in $\mu$G) and on the magnetic field equivalent to the microwave background $B_{\rm CMB} =3.25 (1+z)^2$ (in $\mu$G) as (Murgia et al. 2002):
\begin{equation}
\tau_{\rm syn} = 1610 {\frac{B^{0.5}}{B^2 + B^2_{\rm CMB}}}{\frac{1}{[\nu_{\rm br}(1+z)]^{1/2}}}
\label{b}
\end{equation}
However, in the case of emission from relativistic jets, like most of sources in our samples, the break frequency is up-shifted by Doppler effects. 

\begin{table}
\caption{Mean and interquartile ranges for spectral indices in various spectral ranges including data obtained for the PACO Bright source sample from ATCA (first four rows) and {\it Planck} (rows 5-9). For each couple of frequencies we included only the number of sources (last column) with a 5 sigma detection at both frequencies.}\label{tab:cc}
\begin{tabular}{ccccc}
\hline
Freq.    & Median & first  & third & \#\\
Range [GHz]    &       & quart.        &quart.&  \\
\hline
5-9         &  0.05  &  -0.012  &  0.22 &165\\
9-18        & -0.12 &  -0.32  &  0.05 &166\\
18-33       & -0.32  & -0.50  &  -0.08 &144\\
33-39    & -0.38  & -0.64  &  -0.17 &143\\
44.1-70.4  & -0.40 &  -0.60  &  -0.24 &47\\
70.4-100    & -0.65  & -0.90  &  -0.42& 53 \\
100-143   & -0.67  & -0.94  &  -0.45 & 110\\
143-217    & -0.57  & -0.83  &  -0.45& 122\\
217-353     & -0.61 & -0.80  &  -0.23& 37\\
\hline
\end{tabular}
\end{table}

\begin{figure}
\begin{center}
\includegraphics[width=6cm, angle=90]{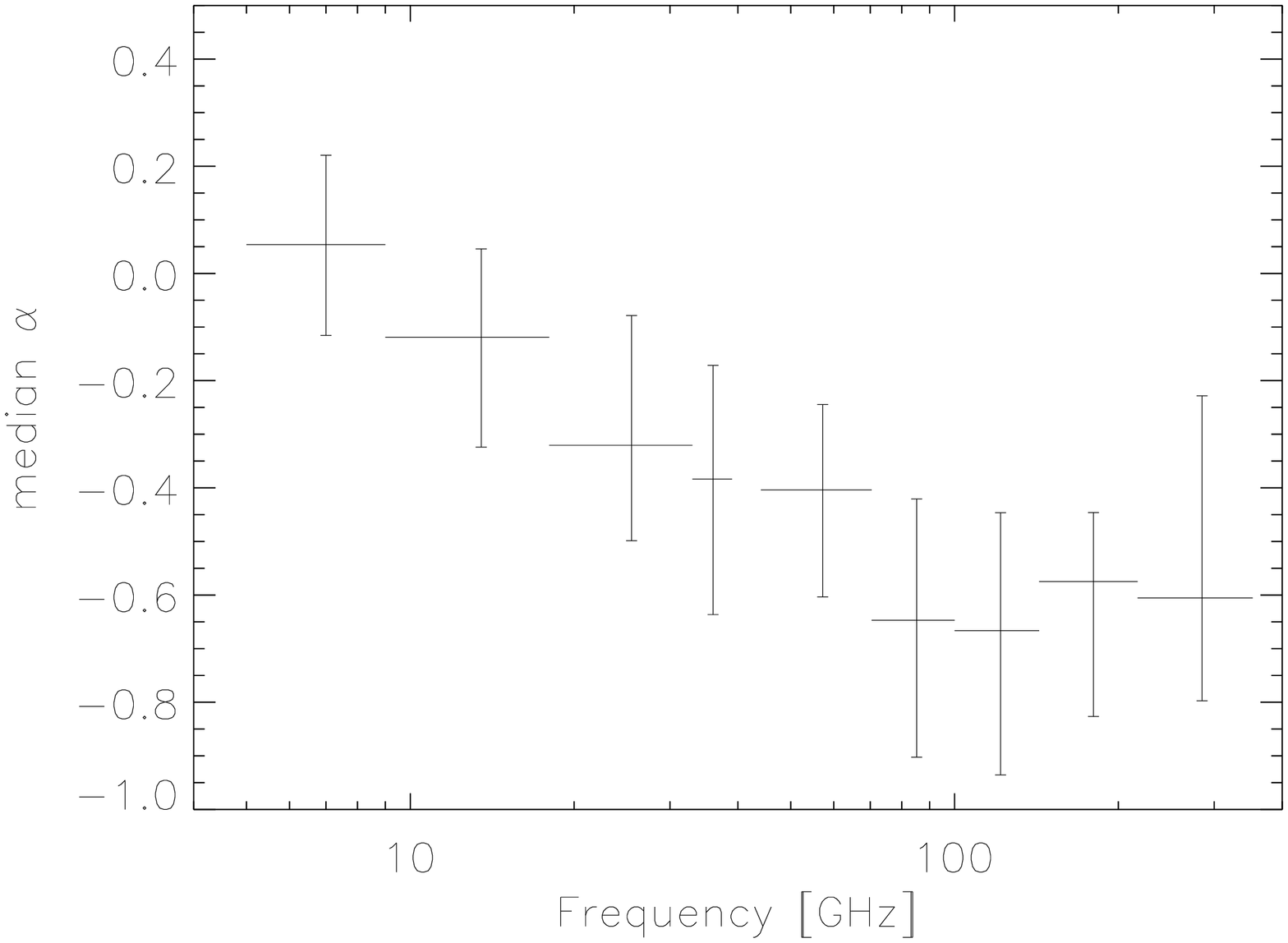}
\caption{Median spectral indices in various spectral ranges including data obtained for the PACO Bright source sample from ATCA (below 30\,GHz) and {\it Planck} (above 30\,GHz). For each couple of frequencies we included only the number of sources with a 5 $\sigma$ detection at both frequencies (see Table \ref{tab:cc}, last column). Error bars are given by the first and third quartile of the spectral indices distribution. Overall, the mean spectral index is peaking at low frequencies and steepening, but no break is observable in the {\it Planck} frequency ranges. Rising of dust contamination is barely seen at very high frequencies. See the text for details.}\label{fig:median_alpha}
\end{center}
\end{figure}

We investigated the variation of the spectral indices with frequency for the PACO Bright Source sample, for which we have a larger probability to find detections at {\it Planck} frequencies. Table~\ref{tab:cc} and Fig.~\ref{fig:median_alpha} summarize the median values of spectral indices in the 4--217\,GHz frequency range. No break is clearly visible at high frequencies. The spectral indices in the ranges 70--100\,GHz and 100--150\,GHz (-0.65 and -0.67 respectively) have 98.3\,\% probability to be drawn from the same distribution according to a Kolmogorov-Smirnov test. At even higher frequencies spectral indices seem to increase, probably as a consequence of the arising of Galactic dust contamination in the vicinity of the sources.

The lack of a spectral break up to mm wavelengths may be taken as an indication that sources must be young, although a quantitative estimate requires the knowledge of the magnetic field intensity. However, even though the bright--sample sources that have detections at {\it Planck} frequencies constitute a large enough sample to infer statistically significant conclusions, we should keep in mind that the high flux density selection might bias towards sources with flatter spectra up to very high frequencies and the low sensitivity of {\it Planck} detections might preferentially select sources that do not have any break. Hence to draw a conclusion deeper observations at {\it Planck} frequencies are needed for a fainter flux-density selected sample. Also, there are several issues related to ageing calculations based on the spectral break \citep{Rudnick2002}. In particular, non-standard electron acceleration mechanisms can effectively counter the electron ageing effect \citep{Kardashev1962}. Indications of such non-standard mechanisms have been pointed out by \citet{PlanckCollaborationXV2011} based on the analysis of the radio continuum spectra of 104 extragalactic radio sources using measurements reported in the Planck Early Release Compact Source Catalogue \citep[ERCSC;][]{PlanckCollaborationVII2011} complemented with simultaneous multifrequency data.

Sources classified as steep-spectrum at low radio frequencies are a minor constituent of our sample and therefore their statistics is poor. Nevertheless it is worth noticing that also in this case there is no evidence of a high frequency break. Also they do not show evidence of the emergence of a compact nucleus at wavelengths of a few cm.


\section{Conclusions} \label{sec:conclusions}

The PACO project consists in observations with ATCA in three pairs of 2\,GHz bands (5.5 and 9\,GHz, 18 and 24\,GHz, 33 and 39\,GHz) of several complete samples of sources  drawn from the AT20G catalogue. As found by Massardi et al. (2011a), the catalogue is dominated by flat-spectrum sources, particularly at the bright fluxes of interest here.  The observations were carried out in 65 epochs between July 2009 and August 2010. For each source at least one observation was made within 10 days from the {\it Planck} satellite observations in any of the LFI channels (30, 44, 70\,GHz).

We presented the data for all epochs of observations for the 464 sources extracted from the AT20G survey; 344 of them form a set of partially overlapping complete samples. The remaining sources have been selected as candidate flaring sources during the first two {\it Planck} epochs. The comparison with the AT20G multifrequency data identified a variation of the spectral behaviour of most flat/inverted spectrum sources in the PACO epochs with respect to the selection epoch, consistent with a blazar classification.

The comparison with the {\it Planck} PCCS catalogue allowed us to describe the population behaviour over a wide spectral range, from 5 to 217\,GHz. The vast majority ($\simeq 91\,\%$) of our sources have a remarkably smooth spectrum, well described by a double power law over the full range. This suggests a single emitting region with an appropriate density profile, at variance with the notion that ``flat'' spectra result from the superposition of the emissions from a number of different compact regions, self absorbed up to different frequencies.

Most of the objects show a spectral steepening above $\simeq 30\,$GHz, consistent with synchrotron emission becoming optically thin. Thus, the spectral slope at mm wavelengths no longer allows us to discriminate between compact (flat-spectrum) and extended (steep-spectrum) radio sources, as done at cm wavelengths. Thus, the classical dichotomy between flat-spectrum/compact and steep-spectrum/extended radio sources, well established at cm wavelengths, breaks down at mm wavelengths.

The mm-wave spectra do not show indications of the spectral break expected as the effect of `electron ageing', suggesting young source ages. It also does not show any indication of dust-like rising of the spectra that could be associated to stellar formation in the host galaxies.

\section*{Acknowledgments}
We thank Jamie Stevens (CSIRO, Australia) and Bruce Partridge (Haverford College, USA) for the useful discussions about the new model for our flux density calibrator.
We thank the anonymous referee for useful comments.
MM and GDZ acknowledge financial support from ASI/INAF Agreement 2014-024-R.0 for the {\it Planck} LFI activity of Phase E2.
AB acknowledges support from the European Research Council under the EC FP7 grant number 280127.
MLC acknowledges the Spanish MINECO Consolider-Ingenio 2010 CSD2010-00064.
MM and VG acknowledge financial support by the Italian Ministero dell\'\,Istruzione, Universit\`a e Ricerca through the grant Progetti Premiali 2012 -- iALMA (CUP C52I13000140001).

We thank the staff at the Australia Telescope Compact Array site, Narrabri (NSW), for the valuable support they provide in running the telescope and in data reduction. The Australia Telescope Compact Array is part of the Australia Telescope which is funded by the Commonwealth of Australia for operation as a National Facility managed by CSIRO.

\bsp

\label{lastpage}


\begin{thebibliography}{}

\bibitem[\protect\citeauthoryear{Abdo et al.}{2010}]{Abdo2010} Abdo A.~A., et al., 2010, ApJ, 716, 30

\bibitem[\protect\citeauthoryear{Blandford {\&} K{\"o}nigl}{1979}]{BlandfordKonigl1979} Blandford R.~D., K{\"o}nigl A., 1979, ApJ, 232, 34

\bibitem[\protect\citeauthoryear{Bonaldi et al.}{2013}]{Bonaldi2013} Bonaldi A., Bonavera L., Massardi M., De Zotti G., 2013, MNRAS, 428, 1845

\bibitem[\protect\citeauthoryear{Bonavera et al.}{2011}]{Bonavera2011} Bonavera L., Massardi M., Bonaldi A., Gonz{\'a}lez-Nuevo J., de Zotti G., Ekers R.~D., 2011, MNRAS, 416, 559

\bibitem[\protect\citeauthoryear{Chhetri et al.}{2012}]{Chhetri2012} Chhetri R., Ekers R.~D., Mahony E.~K., Jones P.~A., Massardi M., Ricci R., Sadler E.~M., 2012, MNRAS, 422, 2274

\bibitem[\protect\citeauthoryear{Ciaramella et al.}{2004}]{2004A&A...419..485C} Ciaramella A., et al., 2004, A\&A, 419, 485

\bibitem[\protect\citeauthoryear{de Zotti et al.}{1999}]{1999AIPC..476..204D} de Zotti G., Toffolatti L., Arg{\"u}eso F., Davies R.~D., Mazzotta P., Partridge R.~B., Smoot G.~F., Vittorio N., 1999, AIPC, 476, 204

\bibitem[\protect\citeauthoryear{de Zotti et al.}{2010}]{DeZotti2010} de Zotti G., Massardi M., Negrello M., Wall J., 2010, A\&ARv, 18, 1

\bibitem[\protect\citeauthoryear{Eddington}{1913}]{1913MNRAS..73..359E} Eddington A.~S., 1913, MNRAS, 73, 359


\bibitem[\protect\citeauthoryear{Hogg
\& Turner}{1998}]{1998PASP..110..727H} Hogg D.~W., Turner E.~L., 1998, PASP, 110, 727

\bibitem[\protect\citeauthoryear{Hogg}{2001}]{2001AJ....121.1207H} Hogg
D.~W., 2001, AJ, 121, 1207

\bibitem[\protect\citeauthoryear{Impey \& Neugebauer}{1988}]{1988AJ.....95..307I} Impey C.~D., Neugebauer G., 1988, AJ, 95, 307

\bibitem[\protect\citeauthoryear{Kardashev}{1962}]{Kardashev1962} Kardashev N.~S., 1962, SvA, 6, 317


\bibitem[\protect\citeauthoryear{Leach et al.}{2008}]{Leach2008} Leach S.~M., et al., 2008, A\&A, 491, 597

\bibitem[\protect\citeauthoryear{Lister et al.}{2009}]{Lister2009} Lister M.~L., et al., 2009, AJ, 137, 3718

\bibitem[\protect\citeauthoryear{Lu et al.}{2012}]{Lu2012} Lu R.-S., Shen Z.-Q., Krichbaum T.~P., Iguchi S., Lee S.-S., Zensus J.~A., 2012, A\&A, 544, A89

\bibitem[\protect\citeauthoryear{Mahony et al.}{2011}]{2011MNRAS.417.2651M} Mahony E.~K., et al., 2011, MNRAS, 417, 2651

\bibitem[\protect\citeauthoryear{Marscher}{1980}]{Marscher1980} Marscher A.~P., 1980, Natur, 288, 12

\bibitem[\protect\citeauthoryear{Massaro et al.}{2009}]{Massaro2009} Massaro E., Giommi P., Leto C., Marchegiani P., Maselli A., Perri M., Piranomonte S., Sclavi S., 2009, A\&A, 495, 691

\bibitem[\protect\citeauthoryear{Massardi \& Burigana}{2010}]{2010NewA...15..678M} Massardi M., Burigana C., 2010, NewA, 15, 678

\bibitem[\protect\citeauthoryear{Massardi et al.}{2008}]{Massardi2008} Massardi M., et al., 2008, MNRAS, 384, 775

\bibitem[\protect\citeauthoryear{Massardi et al.}{2011a}]{Massardi2011a} Massardi M., et al., 2011a, MNRAS, 412, 318 

\bibitem[\protect\citeauthoryear{Massardi et al.}{2011b}]{Massardi2011b} Massardi M., Bonaldi A., Bonavera L., L{\'o}pez-Caniego M., de Zotti G., Ekers R.~D., 2011b, MNRAS, 415, 1597 


\bibitem[\protect\citeauthoryear{Murgia et al.}{2002}]{Murgia2002} Murgia M., Fanti C., Fanti R., Gregorini L., Klein U., Mack K.-H., Vigotti M., 2002, NewAR, 46, 307

\bibitem[\protect\citeauthoryear{Murphy et al.}{2010}]{Murphy2010} Murphy T., et al., 2010, MNRAS, 402, 2403

\bibitem[\protect\citeauthoryear{Partridge et
al.}{2015}]{2015arXiv150602892P} Partridge B., L{\'o}pez-Caniego M., Perley
R.~A., Stevens J., Butler B.~J., Rocha G., Walter B., Zacchei A., 2015,
arXiv, arXiv:1506.02892


\bibitem[\protect\citeauthoryear{Planck Collaboration I}{2011}]{PlanckCollaborationI2011} Planck Collaboration I, 2011, A\&A, 536, A1

\bibitem[\protect\citeauthoryear{Planck Collaboration XV}{2011}]{PlanckCollaborationXV2011} Planck Collaboration XV, 2011, A\&A, 536, A15


\bibitem[\protect\citeauthoryear{Planck Collaboration VII}{2011}]{PlanckCollaborationVII2011} Planck Collaboration VII, 2011, A\&A, 536, A7 


\bibitem[\protect\citeauthoryear{Planck Collaboration XXVIII}{2014}]{PCCS} Planck Collaboration XXVIII, 2014, A\&A, 571, A28 




\bibitem[\protect\citeauthoryear{Rudnick}{2002}]{Rudnick2002} Rudnick L., 2002, NewAR, 46, 95

\bibitem[\protect\citeauthoryear{{Sault} et~al.}{1995}]{1995ASPC...77..433S}
{Sault} R. J.,  {Teuben} P. J., {Wright} M. C. H.,  1995, in {R.~A.~Shaw,
  H.~E.~Payne, \& J.~J.~E.~Hayes} ed., Astronomical Data Analysis Software and
  Systems IV Vol.~77 of Astronomical Society of the Pacific Conference Series,
  {A Retrospective View of MIRIAD}. pp 433

\bibitem[\protect\citeauthoryear{Toffolatti et
al.}{1998}]{1998MNRAS.297..117T} Toffolatti L., Argueso Gomez F., de Zotti
G., Mazzei P., Franceschini A., Danese L., Burigana C., 1998, MNRAS, 297,
117

\bibitem[\protect\citeauthoryear{Toffolatti et
al.}{1999}]{1999ASPC..181..153T} Toffolatti L., De Zotti G., Arg{\"u}eso
F., Burigana C., 1999, ASPC, 181, 153


\bibitem[\protect\citeauthoryear{Toffolatti et
al.}{2013}]{2013arXiv1302.3355T} Toffolatti L., Burigana C., Argueso F.,
Diego J.~M., 2013, arXiv, arXiv:1302.3355


\bibitem[\protect\citeauthoryear{Wilson et al.}{2011}]{2011MNRAS.416..832W}
Wilson W.~E., et al., 2011, MNRAS, 416, 832

\end{thebibliography}
\end{document}